\documentclass[onecolumn, 12pt, journal]{IEEEtran}
\usepackage{setspace}
\usepackage{cite}
\usepackage{graphicx,balance}

\usepackage[cmex10]{amsmath}
\usepackage{amsfonts}
\usepackage{subfigure}
\usepackage{longtable,balance}
\usepackage{array}
\usepackage{multirow}
\usepackage{url}
\usepackage{color}
\usepackage{threeparttable}
\usepackage{rotating}

\makeatletter

\newcommand{\Rmnum}[1]{\expandafter\@slowromancap\romannumeral #1@}
\makeatother
\usepackage{algorithmic}
\usepackage{algorithm}
\begin{document}
\title{When Full Duplex Wireless Meets Non-Orthogonal Multiple Access: Opportunities and Challenges}
\author{Xianhao Chen,
        Gang Liu,~\IEEEmembership{Member,~IEEE,}
        Zheng Ma,~\IEEEmembership{Member,~IEEE,}\\
        Xi Zhang,~\IEEEmembership{Fellow,~IEEE,}
        Pingzhi Fan,~\IEEEmembership{Fellow,~IEEE,}\\
        Shanzhi Chen,~\IEEEmembership{Senior Member,~IEEE,}
        and F. Richard Yu,~\IEEEmembership{Fellow,~IEEE}

\thanks{The work of X. Chen and G. Liu was jointly supported by NSFC Project (No. 61601377), Sichuan Science and Technology Program (No. 2019YJ0248) and the open research fund of National Mobile Communications Research Laboratory, Southeast University (No. 2019D05). The work of Z. Ma was supported by NSFC Project (No.U1709219), Marie Curie Fellowship (No. 792406), and NSFC China-Swedish project (No. 6161101297). The work of P. Fan was supported by NSFC Project (No.61731017).}
 \thanks{X. Chen, G. Liu and P. Fan are with the Key Lab of Information Coding and Transmission, Southwest Jiaotong University, Chengdu, 610031, China. G. Liu is also with National Mobile Communications Research Laboratory, Southeast University (Corresponding author: gangliu@swjtu.edu.cn).}
 \thanks{Z. Ma is with the Department of Information Science and Engineering, KTH Royal Institute of Technology, Stockholm, Sweden, and also with the Key Lab of Information Coding and Transmission, Southwest Jiaotong University, Chengdu, 610031, China.}
 \thanks{X. Zhang is with the Networking and Information Systems Laboratory, Department of Electrical
and Computer Engineering, Texas A\&M University, College Station, TX 77843, USA.}
\thanks{S. Chen is with the State Key Laboratory of Wireless Mobile Communications, China Academy of Telecommunication Technology, Beijing 100191, China, and also with the State Key Laboratory of Networking and Switching Technology, Beijing University of Posts and Telecommunications, Beijing 100876, China.}
\thanks{F. Richard Yu is with the Department of Systems and Computer Engineering, Carleton University, Ottawa, ON, Canada.}}
\maketitle
\begin{spacing}{2}
\begin{abstract}
Non-orthogonal multiple access (NOMA) is a promising radio access technology for the 5G wireless systems. The core of NOMA is to support multiple users in the same resource block via power or code domain multiplexing, which provides great enhancement in spectrum efficiency and connectivity. Meanwhile, with the recent advance in self-interference (SI) cancelation techniques, full duplex (FD) wireless communication has become a feasible technology enabling radios to receive and transmit simultaneously. This article aims to investigate the combination of these two emerging technologies. At first, several typical scenarios and protocols are presented to illustrate the application of FD technique in NOMA systems. Then, a novel NOMA system with FD base stations (BSs) based on centralized radio access networks (C-RAN) is proposed. Furthermore, power allocation policies are discussed for the proposed scheme, and simulation results are provided to demonstrate its superiority. Finally, challenges and research opportunities of FD NOMA systems are also identified to stimulate the future research.
\end{abstract}
\IEEEpeerreviewmaketitle

\section{Introduction\label{sect: introduction}}
As a promising multiple access technology in the upcoming 5G wireless networks, non-orthogonal multiple access (NOMA) has recently attracted intensive research attention. In contrast to orthogonal multiple access (OMA) that serves users in different orthogonal resource blocks, NOMA is capable of serving multiple users with different quality of service (QoS) requirements using the same time-frequency resource by either power-domain or code-domain multiplexing, therefore offering a number of advantages, including improved spectrum efficiency, reduced end-to-end latency, and massive connectivity\cite{Ding2017A}.

On the other hand, full duplex (FD) wireless is another emerging technique for future wireless networks\cite{LYHLL15}. In the past, a long-held belief is that wireless radios can only operate in half-duplex (HD) mode, meaning that they can only transmit and receive either over different time slots or over different frequency bands. However, with recent advances in self-interference (SI) cancellation technologies, the feasibility of FD wireless has been demonstrated, which allows radios to receive and transmit on the same frequency band simultaneously. Moreover, the architectural progression towards short-range systems, such as small-cell systems and WiFi, where the cell-edge path loss is less than that in traditional cellular systems, making the SI reduction more manageable. These changes have recently sparked a significant interest on FD.

Inspired by the aforementioned potential benefits of NOMA and FD technique, it is natural to investigate the promising applications of FD technique in the NOMA systems for further performance improvement. For example, the spectral efficiency of NOMA systems can be further improved by letting the base station (BS) execute the uplink (UL) and downlink (DL) transmissions with FD mode. Moreover, FD relaying (FDR) can potentially double the spectral efficiency of cooperative NOMA systems. Motivated by these observations, this article aims to provide a survey of the recent works on how to combine NOMA with FD, and propose several novel scenarios incorporating these two technologies to stimulate future research. First, we briefly introduce the basic concepts of NOMA and FD techniques. Then, several typical transmission protocols of FD NOMA systems are presented to elaborate how to integrate the FD technique into NOMA. In addition, with the aim of exploiting the merits of both FD and NOMA as well as mitigating the resulted inter-cell interference in the multi-cell NOMA systems, we propose a multi-cell FD NOMA systems based on C-RAN architecture, where the BS operates DL and UL NOMA in FD mode. Our simulation results reveal that the C-RAN scheme can achieve a significantly higher system performance than the traditional schemes. Finally, we identify the challenges and future research opportunities of FD NOMA systems.

\section{Overview of Non-Orthogonal Multiple Access and FD Wireless\label{sect: overview}}
\subsection{Non-Orthogonal Multiple Access Techniques}
Unlike conventional OMA, NOMA serves a number of users in non-orthogonal resource block, which can broadly be divided into power-domain NOMA, code-domain NOMA, and other NOMA schemes, such as interleaving or scrambling NOMA, spatial division multiple access (SDMA), bit division multiplexing (BDM), compressive sensing (CS)-based NOMA and so on\cite{Ding2017A,dai2018survey}. Since power-domain and code-domain NOMA have received most research attentions, we only focus on power-domain and code-domain NOMA due to the limitation of space in this article.
\begin{figure*}[!t]
\centering
\includegraphics [width=4in]{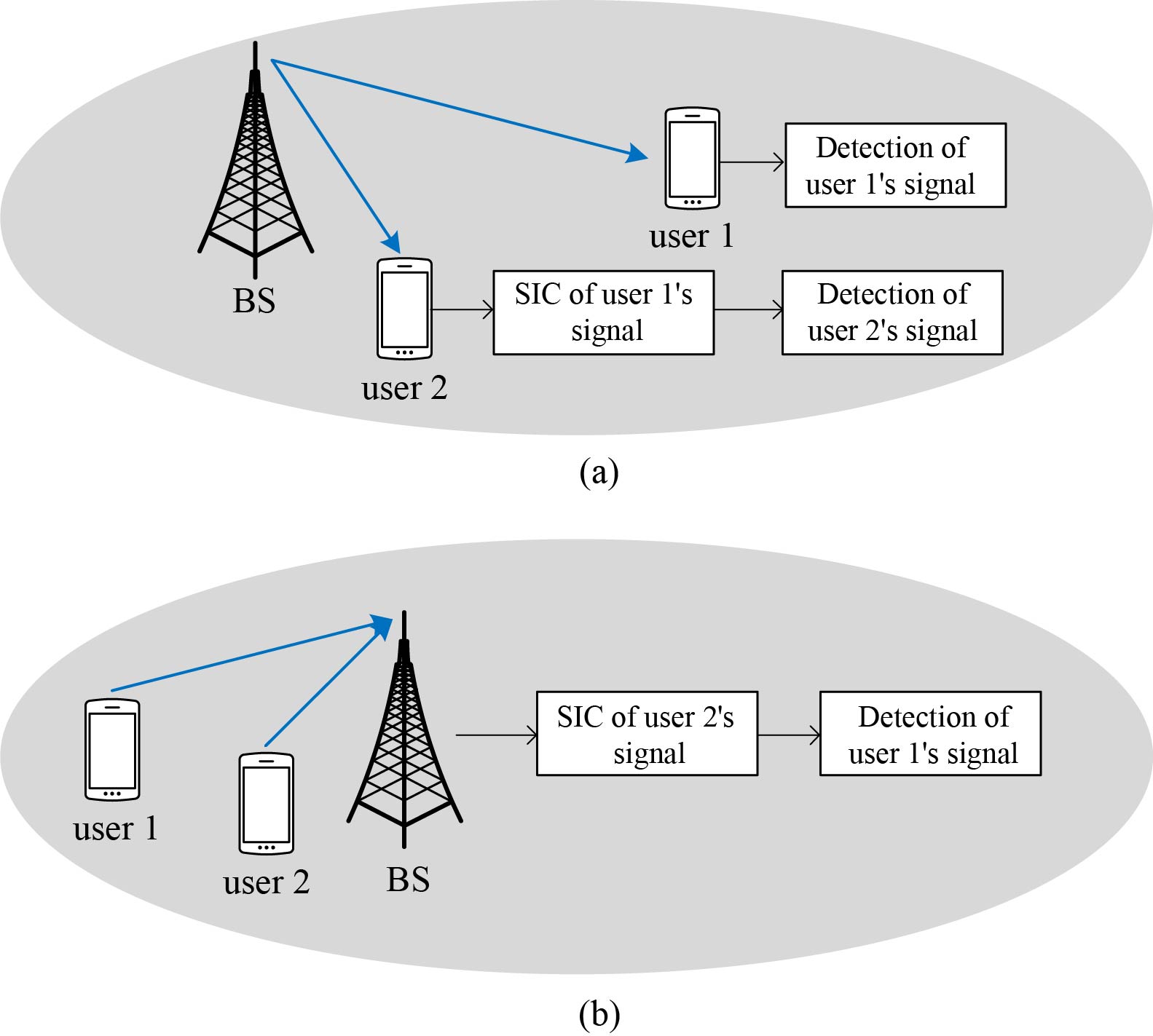}
\caption{Illustration of power-domain NOMA principles. User 2 is with better channel conditions, and user 1 is with poorer channel conditions. (a) DL power-domain NOMA transmission. (b) UL power-domain NOMA transmission.}
\label{fig: DLULNOMA}
\end{figure*}
\subsubsection{Power-domain NOMA}
Power-domain NOMA superposes multiple users in power domain and exploits the channel gain difference between multiplexed users\cite{Ding2017A}. At the transmitter side, signals from various users are superposed and the resulting signal is then transmitted over the same time-frequency resources. By allocating different power levels to different users, different users can be distinguished. At the receiver sides, successive interference cancellation (SIC) is employed to detect the desired signals. The basic idea of SIC is that user signals are successively decoded. Take two-user NOMA system shown in Fig. \ref{fig: DLULNOMA} as an example. In DL NOMA, SIC operation is carried out at a strong user for canceling the weak user's interference as in Fig. \ref{fig: DLULNOMA}(a). By contrast, in UL NOMA, SIC is carried out at the BS to decode and substract the strong user's signal first as shown in Fig. \ref{fig: DLULNOMA}(b).

\subsubsection{Code-domain NOMA}
Typical examples of code-domain NOMA are pattern division multiple access (PDMA) and sparse code multiple access (SCMA), which are based on the idea that one user's information is spread over multiple subcarriers \cite{chen2017pattern,Ding2017A}. According to code-domain NOMA principle, each user is identified by a codebook containing multiple codewords. At the transmitter, bit streams of each user are directly mapped to different sparse codewords of the corresponding codebook. In SCMA systems, the number of subcarriers assigned to each user is smaller than the total number of subcarriers for guaranteeing a manageable system complexity. In PDMA systems, the number of subcarriers assigned to one user is not necessarily much smaller than the total number of subcarriers. Since one user's messages at different subcarriers are jointly encoded, both PDMA and SCMA require joint decoding at the receiver, where the message passing algorithm (MPA) should be adopted to ensure low complexity.
\subsection{Full Duplex Wireless Communications}

\begin{figure*}[!t]
\centering
\includegraphics [width=6in]{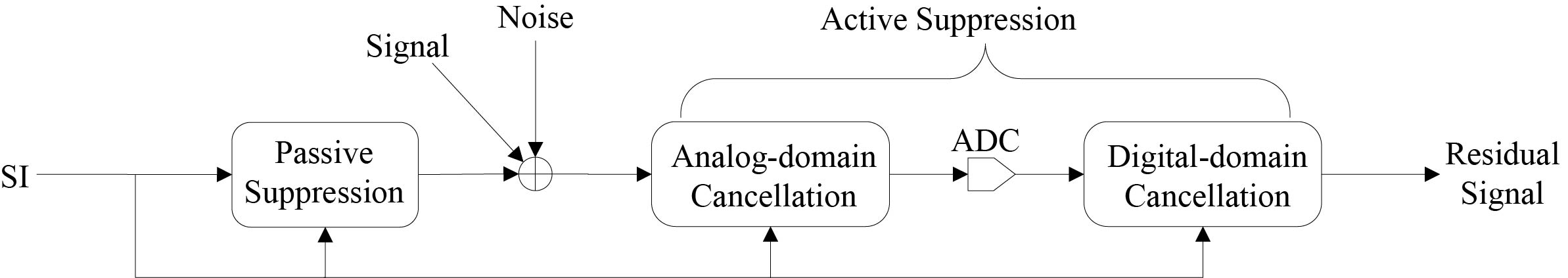}
\caption{Illustration of SI cancellation technologies for full duplex wireless communications.}
\label{fig: FD}
\end{figure*}

FD wireless technology enables wireless terminals to transmit and receive simultaneously over the same frequency band, which potentially doubles the system spectral efficiency. However, the transmitted signals can loop back to the receive antennas and thus cause the SI. To enable FD wireless, the SI must be mitigated sufficiently via a combination of SI cancellation technologies\cite{LYHLL15}. In general, the SI cancellation technologies can be categorized as passive suppression and active suppression as shown in Fig. \ref{fig: FD}.

Passive suppression is defined as the attenuation of the SI signal contributed by the path-loss effect due to the physical separation/isolation between the transmitter and receiver of the same node. For example, for communication systems with separate antenna deployment, path loss can be leveraged to suppress the SI by increasing the physical distance between transmit and receive antennas, or exploiting the surrounding obstacles (e.g., buildings, tunnels, and shielding plates) to block the SI propagating directly from the transmit chain to the receive chain (i.e., direct paths).

Active suppression eliminates SI by subtracting a processed copy of the transmitted signal from the received signal, which can be further categorized into two stages: analog-domain cancellation and digital-domain cancellation. Analog-domain SI cancellation is to cancel the SI in the analog receive-chain circuitry by subtracting a predicted copy of SI from the received signal prior to the digitization. Digital-domain cancellation is to cancel the SI after the analog-to-digital conversion (ADC) as the last line of defense against SI. Basically, since the transmitted signals are typically known at the device, digital-domain cancellation technologies can easily exploit the knowledge of the transmitted signal to subtract it after the received signal has been quantized by the ADC.
\section{Typical Transmission Protocols of FD NOMA\label{sect: protocol}}
In this section, we present several transmission protocols to elaborate the typical applications of FD technique in NOMA systems.
\subsection{Dedicated Relay Cooperation}
\begin{figure*}[!t]
\centering
\includegraphics [width=5in]{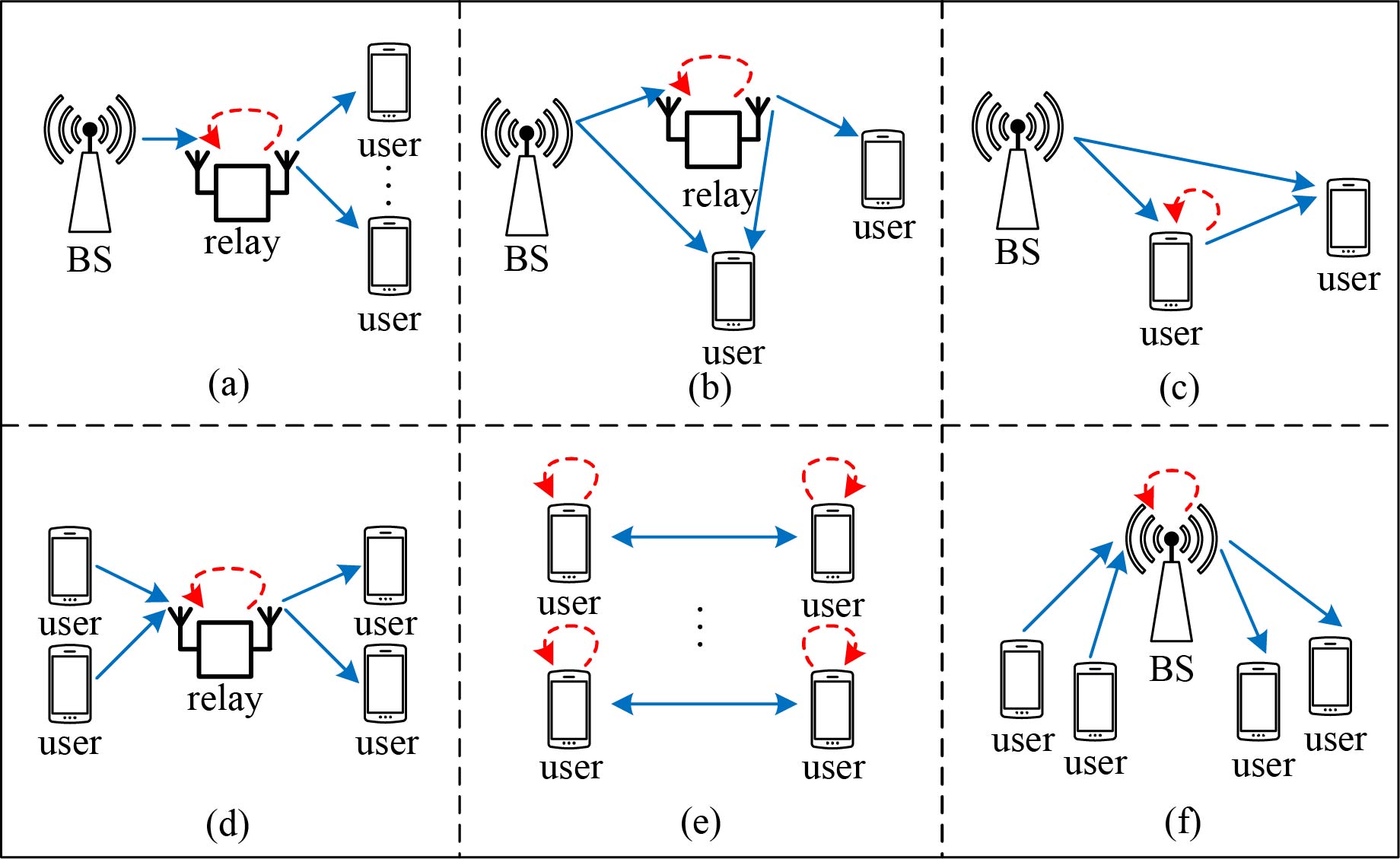}
\caption{The illustration of different FD-NOMA protocols. The blue solid lines represent the transmitted signals, and the red dotted lines represent the SI signals. (a) FD relay-assisted cooperative NOMA scheme. (b) FD relay-assisted cooperative NOMA scheme with coordinated direct and relay transmission. (c) FD user-assisted cooperative NOMA scheme. (d) FD cooperative relay sharing NOMA scheme. (e) FD SCMA/PDMA-based device to device communications. (f) NOMA scheme with full-duplex base station.}
\label{fig: FDRNOMA_models}
\end{figure*}
Relaying technique can provide significantly improved throughput and coverage\cite{LYHLL15}. Thus, it will be promising to integrate FD relaying in NOMA systems. A typical FD relay-assisted cooperative NOMA scheme (FDR-NOMA) is depicted in Fig. \ref{fig: FDRNOMA_models} (a). The BS sends the superposed signal of multiple users to the relay using the NOMA technology. At the same time, the FD relay forwards its received signal in either amplify-and-forward (AF) or decode-and-forward (DF) mode to the users. Finally, each user decodes the superposed signal from the relay by adopting SIC technique.

The authors in \cite{zhong2016non} discussed another typical FDR-NOMA scheme as shown in Fig. \ref{fig: FDRNOMA_models} (b), where the nearby user directly communicates with the BS, while the far user requiring the assistance of the FD relay to communicate with BS. The main benefit of this scheme lies in the fact that the nearby user can decode the far user's data from the BS by adopting SIC strategy, which can be utilized as the side information for cancelling the interference from the relay when the relay forwards the message to the far user.

\subsection{User Cooperation}
In addition to cooperative NOMA scheme with dedicated relay, another form of cooperative NOMA scheme has been investigated as well, in which the nearby users act as relays to assist far users. The user cooperation strategy for NOMA scheme is based on the observation that the the nearby users need to decode the signals of the far users, implying that they can be directly utilized as DF relays to assist the weaker users. The authors in \cite{zhang2016full} investigated an FD user-assisted cooperative NOMA (FDU-NOMA) scheme as shown in Fig. \ref{fig: FDRNOMA_models} (c), in which the BS serves two users simultaneously by using NOMA technology, and the nearby user assists the far user in FD mode by using the message known during the SIC procedure. Maximal ratio combining (MRC) technique can be adopted at the far user to handle the two incoming signals from BS and the nearby user.
\subsection{Cooperative Relay Sharing}
Sharing cooperative relay by multiple users to reduce network deployment expense is an interesting topic of research. The application of NOMA to FD cooperative relay sharing network was studied in \cite{Kader2018Full}, where two source-destination pairs share a dedicated FD relay as in Fig. \ref{fig: FDRNOMA_models} (d). Following the principle of UL NOMA, both sources transmit their symbols to the FD relay. The FD relay decodes these symbols and simultaneously transmits a superimposed signal to the destinations based on the principle of DL NOMA.

\subsection{Device to Device Communications}
D2D communication can largely reduce the end-to-end latency by enabling devices to communicate with each other directly without the assistance of the BS. In order to improve the performance of D2D communication networks, we propose an FD PDMA/SCMA-based D2D communication (FDD-PDMA/SCMA) scheme in Fig. \ref{fig: FDRNOMA_models} (e), where each user transmits the message to its paired user and simultaneously receives the message from its paired user on the same frequency band. For instance, a case where six pairs of users are densely located, the typical factor graph matrix for SCMA is given in \cite{Ding2017A}, and the typical PDMA pattern matrix is given in \cite{chen2017pattern}. At the receiver side, MPA algorithm is performed to jointly decode the desired signal from the paired user. In this way, the FDD-PDMA/SCMA scheme with six user pairs only occupies four subcarriers, thus enhancing spectrum efficiency.

Similarly, by adopting the PDMA/SCMA principle, we can also extend the FDD-PDMA/SCMA scheme into a broadcasting scheme, where each user tends to broadcast its message to other users in the group, and simultaneously receives the broadcasting messages from other users on the same frequency band. In this way, a broadcasting scheme with six users only occupies four subcarriers as well.

\subsection{Full-Duplex Base Station}
Traditionally, the BS operates in HD mode, where DL and UL transmissions occupy orthogonal radio resources, which leads to spectrum underutilization. In fact, the spectral efficiency of traditional NOMA systems can be further improved through enabling the BS to operate in FD mode as in Fig. \ref{fig: FDRNOMA_models} (f) \cite{SNDS17}, which are termed as NOMA systems with FD BS (FDB-NOMA) for simplicity. However, while the FD BS improves the throughput of cellular networks, it brings out additional sources of interference, e.g., the SI, and the intra-cell interference caused on the DL signals by the UL transmission of users.

Fig. \ref{fig: FDRNOMA_models} (f) can also represent an FD DL broadcasting system. In the UL transmission, the UL users transmit their signals to the BS following the uplink NOMA principle, resulting in the SI at the BS, and the interference caused on the DL signals. In the DL transmission, the BS broadcasts the superposed signals to multiple DL users based on NOMA principle. The superposed DL signal can be divided into multiple layers with different transmission power. At the DL user side, each layer is successively decoded from the superposed signal by SIC. In such a way, the receivers with better channel conditions can decode more data to obtain better service, and vice versa, which utilizes the difference in channel conditions among users to improve strong users' QoS, as well as guarantee the QoS of weak users.

\begin{figure*}[!t]
\centering
\includegraphics [width=5in]{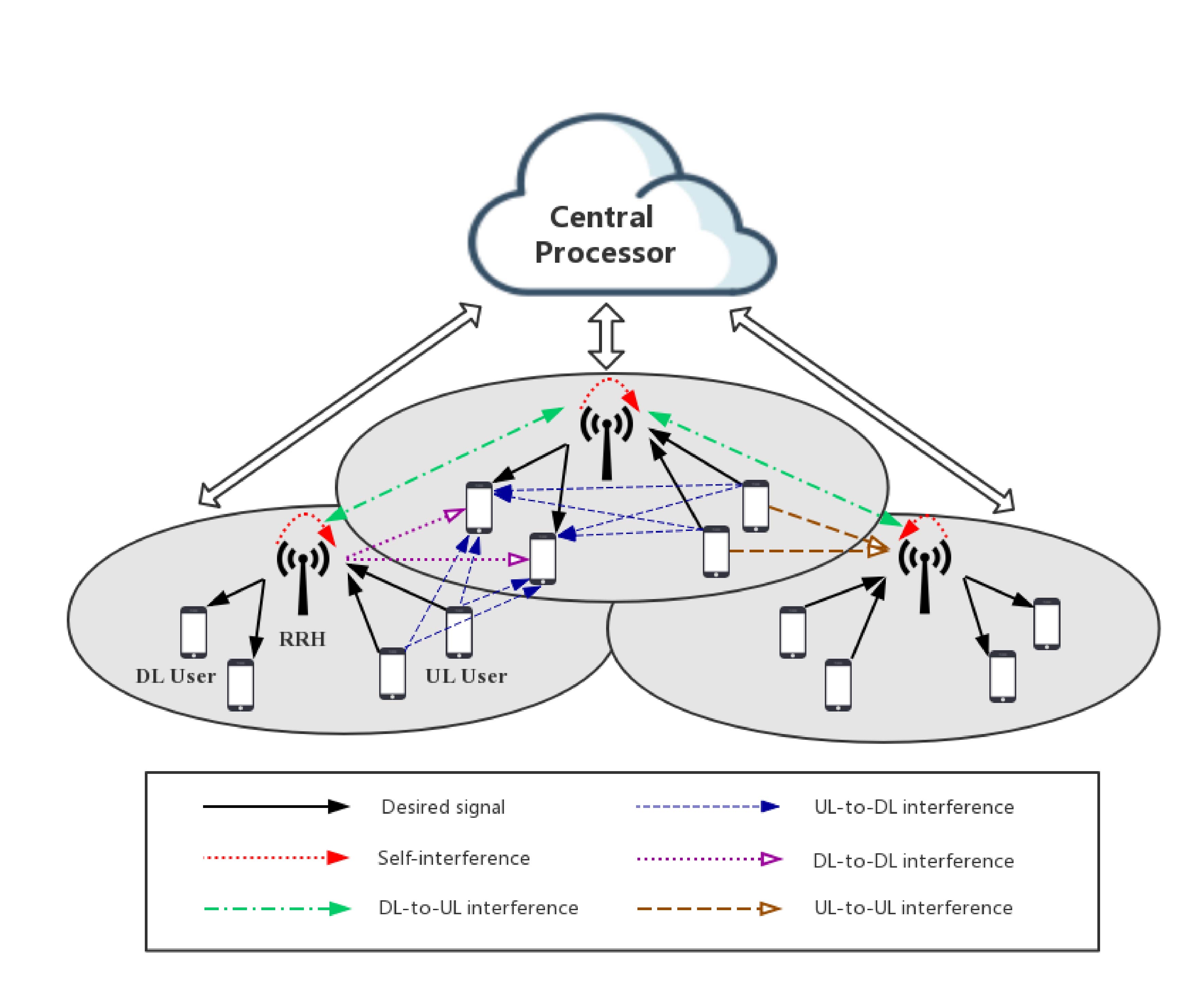}
\caption{Illustration of centralized FD NOMA system.}
\label{fig: CloudBSs}
\end{figure*}

\section{Case Study: Centralized FDB-NOMA System \label{sect: cloud}}
In this section, we propose a centralized multi-cell power-domain FDB-NOMA scheme (C-FDB-NOMA), to manage the intra-cell and inter-cell interferences, followed by the corresponding power allocation schemes and performance evaluations. Since the structures of FD code-domain NOMA systems and FD power-domain NOMA systems are similar, we only take power-domain NOMA as an example to show the basic idea of designing a centralized FD NOMA systems. For the sake of simplicity, we will use ``NOMA" to denote ``power-domain NOMA" in this section.

\subsection{Transmission Scheme of C-FDB-NOMA}
As aforementioned, intra-cell and inter-cell interferences both occur in the multi-cell FDB-NOMA systems, which makes the interferences more difficult to manage. Generally speaking, in addition to the interferences inherently caused by the DL and UL NOMA transmissions, the following interferences should be also considered in the multi-cell FDB-NOMA systems as shown in Fig. \ref{fig: CloudBSs},
\begin{itemize}
\item \textbf{DL-to-UL interference}: The inter-cell interference caused on the UL signals by the DL transmission of BSs from other cells.
\item \textbf{UL-to-DL interference}: The interference caused on the DL signals by the UL transmission of users, which can be either inter-cell or intra-cell interference.
\item \textbf{DL-to-DL interference}: The inter-cell interference caused on the DL signals by the DL transmission of BSs from other cells.
\item \textbf{UL-to-UL interference}: The inter-cell interference caused on the UL signals by the UL transmission of users from other cells.
\item \textbf{Self-interference}: Self-interference at BSs due to the FD operation.
\end{itemize}

Due to the severe interference in the FDB-NOMA systems, it is crucial to develop effective schemes to suppress the interference and satisfy the QoS of users. To this end, the authors in \cite{SNDS17} investigated the joint power allocation and subcarrier allocation of single-cell FDB-NOMA systems, assuming that each subcarrier is allocated to at most two DL users and two UL users. The authors in \cite{elbamby2017resource} studied the joint optimization problem of mode selection, user association and power allocation in a multi-cell NOMA system, with the objective of investigating the benefits of operating in HD or FD, as well as in OMA or NOMA modes, depending on different channel conditions. However, this work assumes that the FD mode and NOMA mode can not be simultaneously employed.

With the aim of exploiting the merits of both FD and NOMA as well as mitigating the interference caused by them, we propose a C-FDB-NOMA system as shown in Fig. \ref{fig: CloudBSs}. In the considered scenario, the baseband processing and medium access control (MAC) functions are moved into a central unit (CU). Consequently, the traditional high-cost BSs can be replaced by cost-effective and power-efficient radio remote heads (RRHs) that only retain radio functionality. This migration of baseband processing is enabled by a network of fronthaul links, such as fiber optics cables or mmwave radio links, that connect each RRH to the CU. In the DL transmission, the data streams are jointly precoded at CU, and are delivered to RRHs via fronthaul links and then to users via wireless links. In the UL transmission, each RRH compresses its received signal for transmission to CU via fronthaul links. Based on the received compressed signals, the CU performs joint decoding of the data streams of all users. The centralized baseband processing enables CU to perform cancellation of the DL-to-UL interference, since the DL signals of RRHs are already known at the CU. From another perspective, since all the BSs are connected to the CU, DL-to-UL interference can be viewed as SI at the CU. Hence, traditional SI cancellation technologies\cite{LYHLL15} in FD networks can be used to cancel the DL-to-UL interferences. Generally speaking, the proposed C-RAN architecture has the following advantages. First, the centralized resource optimization (e.g., power allocation) and signal processing can be realized to cooperatively suppress both the intra-cell and inter-cell interferences. The CU can employ RRHs to collect network information, such as the instantaneous channel state information (CSI) and spectrum information. Based on the collected information, the CU obtains resource allocation strategies accordingly, and then sends decisions to RRHs and users within its coverage to coordinate their operation.  In such a way, the high interference resulting from performing both NOMA and FD can be significantly suppressed. Second, since the C-RAN structure simplifies the implementation of RRHs, RRHs can be widely deployed in close proximity to users in order to enhance the system performance via short-range transmissions and spectrum reuse.
\subsection{Power Allocation Scheme for C-FDB-NOMA}
In order to further mitigate the high interference of multi-cell NOMA system, a well-designed power allocation scheme is needed. For simplicity, let us assume that each user associates to the nearest BS. In order to maximize the sum throughput, the power allocation problem is formulated as follows,
\begin{align}
&\max_{\textbf p} \sum\limits_{d\in \mathcal D} \log_2 (1+\text{SINR}_d (\textbf p)) \label{P1}\\
\text{s.t.}  \ \   &\text{C0}:\ \textbf p \in \mathcal P, \notag\\  &\text{C1}: \  \log_2 (\text{SINR}_d (\textbf p))\geq r_{\text{min},d},\ \ d\in \mathcal D, \notag
\end{align}
where $\mathcal D$ denotes the set of users, $\textbf p$ denotes the collection of power allocation variables, including transmission powers of BSs and UL users. $\text{SINR}_d (\textbf p)$ denotes the signal to interference-and-noise ratio (SINR) of user $d$, which is a function of $\textbf p$. $\mathcal P$ is the non-negative feasible set accounting for the limited power resources, and $r_{\text{min},d}$ denotes the minimum data rate requirement of user $d$.

Due to the non-convexity of the objective function, the formulated optimization problem is non-convex. Fortunately, thanks to the hidden monotonicity property of the considered problem, we can obtain the optimal power allocation strategy by adopting the monotonic optimization method as in \cite{Qian2009MAPEL}. Towards this direction, we first convert the original optimization problem into the following standard monotonic problem,
\begin{align}
\hspace{10mm}  \max_{\textbf z}& \sum\limits_{d\in \mathcal D} \log_2(z_d) \notag\\
&\text{s.t.}  \ \   \textbf z\in \mathcal{Z}, \label{P2}
\end{align}
where $\textbf z=[z_1,...,z_{|\mathcal D|}]=[1+\text{SINR}_1 (\textbf p),...,1+\text{SINR}_d (\textbf p)]$ , and the feasible sets $\mathcal{Z}= \Big \{\textbf z|2^{r_{d,\text{min}}} \leq z_d\leq 1+\text{SINR}_d (\textbf p),  \textbf p\in\mathcal{P}, \forall d\in \mathcal D\Big \}$. It is noted that the variable in problem (\ref{P2}) has been changed to $\textbf z$. In this way, the reformulated problem is converted into a monotonic optimization problem, with its objective monotonically increasing in $\textbf z$. The optimal solution of problem (\ref{P2}) will be achieved at the upper boundary of the feasible region of $\textbf z$, and thus the procedure of finding the optimal solution is reduced to search for the optimality on the boundary of the feasible region. According to the monotonic optimization theory, we can construct a sequence of polyblocks outer approximating the feasible region with an increasing level of accuracy. This procedure of locating the optimality on the feasible boundary is called as polyblock outer approximation algorithm, and the detailed procedure can be found in \cite{Qian2009MAPEL}.

However, the proposed monotonic optimization procedure demands exponential computational complexity, which hence, is not suitable for large user set. For this reason, we adopt this approach as a benchmark to reveal the optimal performance of our C-FDB-NOMA. To solve problem (\ref{P1}) more efficiently, we transform problem (\ref{P1}) into the following form,
\begin{align}
\max_{\textbf p} \sum\limits_{d\in \mathcal D}& \log_2(S_d(\textbf p)+N_0(\textbf p)+I_m)- \sum\limits_{d\in \mathcal D} \log_2(N_0+I_d(\textbf p)) \notag \\
\text{s.t.}  \ \   &\text{C0}:\ \textbf p \in \mathcal P, \notag\\ &\text{C1}: \  \log_2 (\text{SINR}_d (\textbf p))\geq r_{\text{min},d},\ \ d\in \mathcal D, \label{P3}
\end{align}
where $S_d(\textbf p)$ and $I_d(\textbf p)$ are the received signal and interference power of user $d$, and $N_0$ is the noise power. Since the objective function of (\ref{P3}) is a difference of two concave functions, (\ref{P3}) indeed is a standard DC (difference of convex functions) programming problem. Therefore, we can obtain the sub-optimal solution of (\ref{P3}) by applying successive convex approximation (SCA) algorithm, which converges to local optimality in polynomial time\cite{kha2012fast}. Even though SCA algorithm does not pursuit global optimization, it has been empirically shown to often achieve globally optimal solution in many practical applications, including wireless interference networks as demonstrated in\cite{kha2012fast}. Therefore, SCA algorithm will be a good method to solve our problem efficiently.
\subsection{Performance Evaluations for C-FDB-NOMA}
\begin{table}[h]
\centering
\caption{Simulation parameters}\label{tab:tab2}
\begin{tabular}{|l|c|} 
\hline
The radius of network area&300 meters\\
\hline
Pass loss exponent&3.5\\
\hline
SI channel gain&0 dB\\
\hline
SI cancellation coefficient&-110 dB\\
\hline
DL-to-UL interference cancellation coefficient&-110 dB\\
\hline
Maximum transmission power of RRHs&30 dBm\\
\hline
Maximum transmission power of UL users&27 dBm\\
\hline
Error tolerance for optimal and suboptimal algorithms &0.001\\
\hline
\end{tabular}
\end{table}
Let us investigate the performance of the proposed C-FDB-NOMA scheme through simulations with \textit{MATLAB}. An outdoor cellular network with radius of 300 meters is considered, where 2 BSs and, 4 DL users and 4 UL users are distributed uniformly over the network area. The small-scale fading of the channels are modeled as independent and identically distributed Rayleigh fading, while the SI channel gain being set as a constant. To evaluate the impact of SI, we model the SI at BS side as the product of transmission power and the SI channel gain. In other words, the residual SI increases linearly with the transmission power.
Unless specified in each figure, the parameters referred to simulations are summarized as Table \ref{tab:tab2}. The average throughput of the optimal C-FDB-NOMA is obtained through solving Problem (\ref{P1}) with monotonic optimization method, and the average throughput of the suboptimal C-FDB-NOMA is obtained through solving Problem (\ref{P1}) by transforming it into a DC programming problem. The error tolerance for both algorithms is set as 0.001. The results provided below are averaged over different realizations of both node locations and path loss fading.

For benchmarking, we compare the performance of the proposed algorithms with traditional FDB-NOMA scheme, FDB-OMA scheme and HDB-NOMA scheme. The C-RAN architecture is not considered in the traditional FDB-NOMA scheme, which implies that the DL-to-UL interference cannot be cancelled as in our proposed C-FDB-NOMA scheme. In the FDB-OMA scheme, the RRHs perform DL and UL transmissions in FD mode, but serve DL and UL users by allocating orthogonal resources. In the HDB-NOMA scheme, the RRH performs DL and UL transmissions in HD mode, and serves DL and UL users based on NOMA principle. In all of the benchmark schemes, the optimal power allocation schemes are obtained by monotonic optimization method as well, since hidden monotonicity structure of them can be also easily identified.
\begin{figure*}
  \centering
  \subfigure[Average system throughput against the ratio of RRH maximum transmit power to noise power.]{
    \label{subfig: 1}
    \includegraphics[width=3.3in]{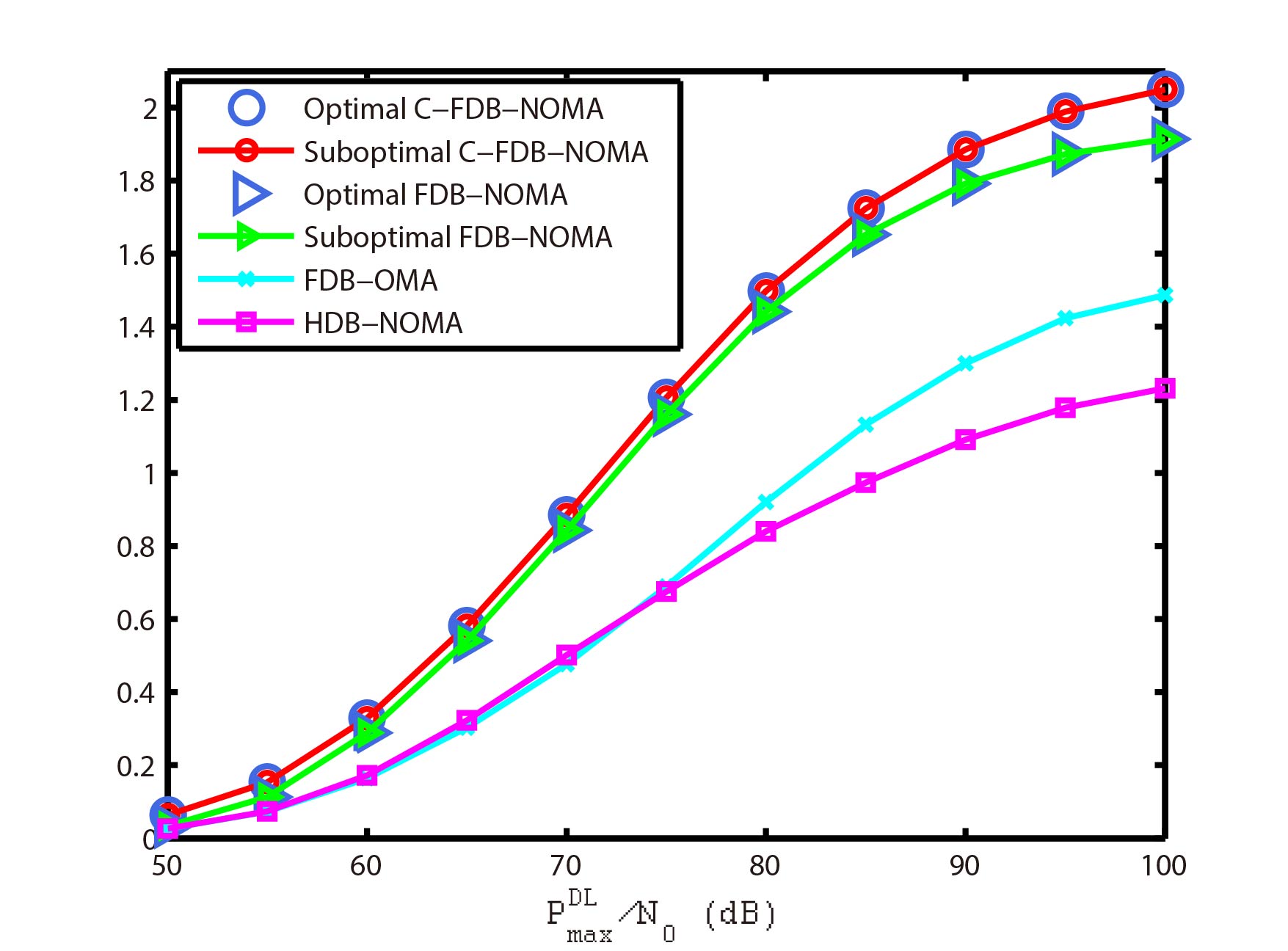}}
  \subfigure[Average system throughput against self-interference cancellation capacity.]{
    \label{subfig: 2}
    \includegraphics[width=3.3in]{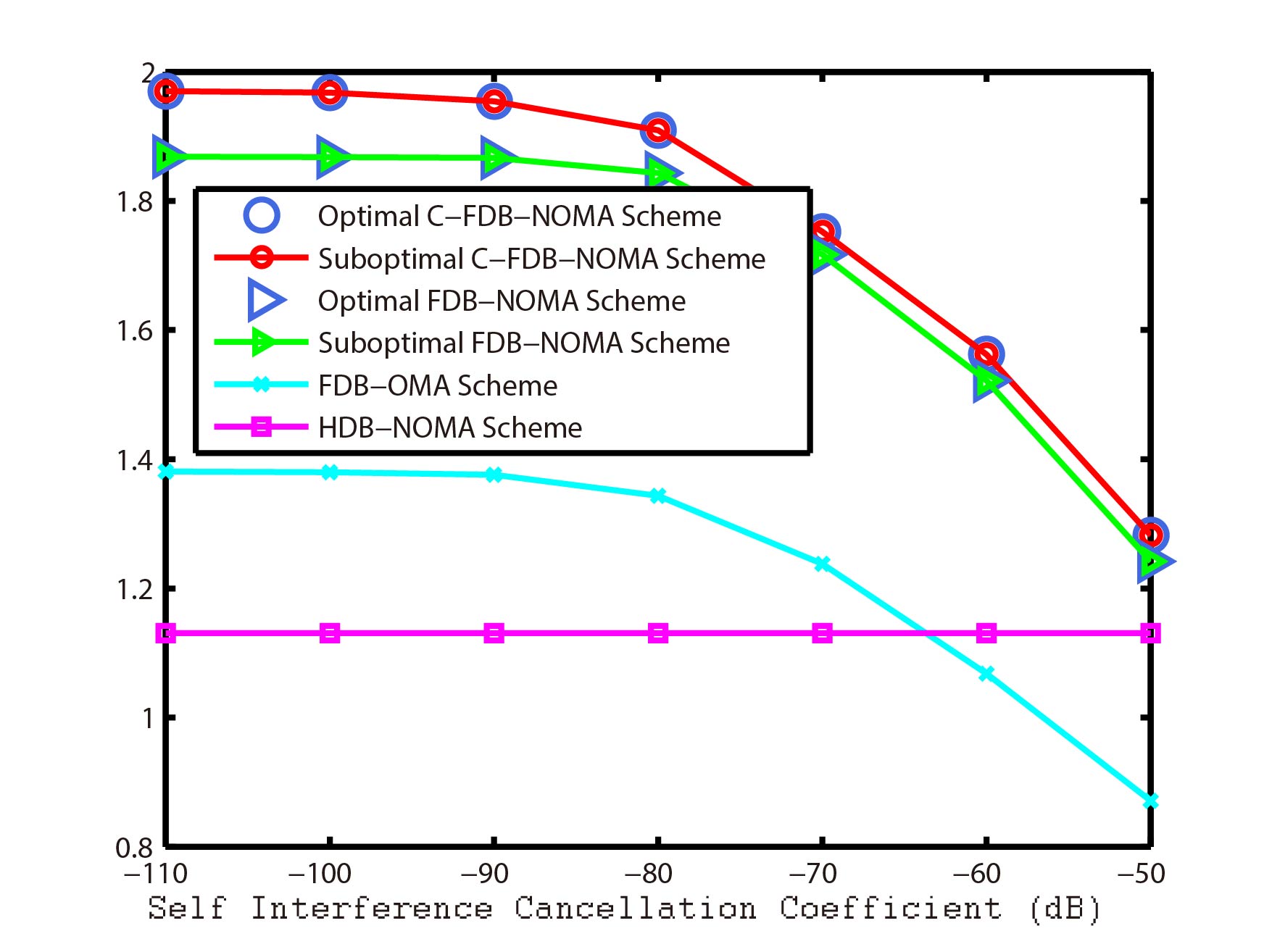}}
  \caption{The average system throughput comparison of different transmission schemes. (a) Average system throughput comparison for varying $P^{\text {DL}}_{\text {max}}/N_0$, with $P^\text {UL}_\text {max}=\frac{1}{2}P^\text {DL}_\text {max}$, where $P^{\text {DL}}_{\text {max}}$ is the maximum DL transmission power at RRH, $P^\text {UL}_\text {max}$ is the maximum UL transmission power of each UL user, and $N_0$ is the noise power at receivers. (b) Average system throughput comparison for varying SI cancellation capacity, with $P^{\text {DL}}_{\text {max}}/N_0=90\text{dB}$, and the minimum user data demand $r_{\text{min}}=0.02 \text{bit/s/Hz}$ for all users.}
  \label{fig: simulation}
\end{figure*}

Fig. \ref{subfig: 1} investigates the average system throughput versus varying $P^{\text {DL}}_{\text {max}}/N_0$ (ratio of RRH maximum transmit power to noise power). As can be seen from Fig. \ref{subfig: 1}, the plots of the suboptimal scheme overlaps with the optimal scheme, verifying that the suboptimal scheme closely approaches the optimal performance. More importantly, it is shown that our proposed schemes provide clearly better performance than the other benchmark schemes. Particularly, the performance gain achieved by our proposed C-FDB-NOMA system over the multi-cell FDB-NOMA system benefits from the fact that the DL-to-UL interference is cancelled in our C-RAN architecture. In addition, the average throughput of each scheme increases fast in low $P^{\text {DL}}_{\text {max}}/N_0$ region, while increasing slowly in high $P^{\text {DL}}_{\text {max}}/N_0$ region. This is because the considered systems are noise-limited systems in low $P^{\text {DL}}_{\text {max}}/N_0$ region, where the noise power is dominant and the interference power is negligible. Even if the total transmit power increases, the increasing interference power does not degrade the performance much, and thus the average throughput will increase fast. In contrast, in the high $P^{\text {DL}}_{\text {max}}/N_0$ region, the considered systems are interference-limited systems. As a consequence, when the total transmit power increases, the interferences (including SI in FD systems) severely degrade the system performance, and therefore the average throughput will increase slowly in this region.

Fig. \ref{subfig: 2} depicts the average performance of different schemes with varying SI cancellation coefficient $\kappa_{\text{SI}}$.
As can be observed, the performance of three FD schemes degrades as the capability of SI cancellation decreases, while the performance of the HDB-NOMA scheme is independent of $\kappa_{\text{SI}}$ due to the HD mode. For this reason, the HD NOMA scheme outperforms the FD-OMA schemes in the case that $\kappa_{\text{SI}}$ is large enough. This phenomenon indicates that the SI cancellation capacity is a key factor influencing the system performance.

\section{Future Research Opportunities and Challenges}
In this section, we will identify the future research opportunities and challenges for FD NOMA systems.
\subsection{Interference and Resource Managements}
Both NOMA and FD techniques impose inevitable interference in networks. The SI induced by FD technique and the inter-user interference induced by NOMA technique makes the resource management problem more complex. Consequently, traditional resource allocation policies cannot be directly applied to FD NOMA networks. On the other hand, well-designed resource allocation policies are pivotal in FD NOMA networks, since the system performance will severely degrade when either the SI or inter-user interference is not properly handled. These observations call for more research efforts on interference and resource management in FD NOMA networks to manage the space-time-frequency resources wisely.

\subsection{FD NOMA with Multiple Antennas}
Multiple-input multiple-output (MIMO) technology is capable of providing significant capacity gain by supporting more spatial streams.
Multiple antennas technique and FD NOMA networks have been combined in a few literature. The authors in \cite{Alsaba2018Full} introduced beamforming and energy harvesting into FD user-assisted cooperative NOMA system. The authors in \cite{Sun2018Robust} study the resource allocation for robust and secure communication in multiple-input single-output (MISO) multicarrier NOMA systems. In the future, multiple antennas can be further merged into various scenarios as described in Section \ref{sect: protocol}.
\subsection{FD NOMA with Caching and Multi-Access Edge Computing}
In the context of wireless caching, popular contents can be proactively predicted and cached in the local content servers (e.g., BS) before they are requested, which facilitates high data rate communications. Although some literature such as \cite{Ding2018NOMACaching} has combined NOMA with caching, as far as we know, few works considered caching technique in the context of FD NOMA system so far. Actually, FD and NOMA techniques can be adopted in wireless caching networks in many ways. For example, when required files are not stored at the content servers, the content pushing and content delivery phases can be executed simultaneously by enabling the content servers to operate in FD mode.

Moreover, multi-access edge computing (MEC) \cite{porambage2018survey} is an emerging technology enabling cloud-computing capabilities in the network edges for providing the delay-sensitive services. By integrating FD technique at the MEC server, the MEC system performs DL and UL transmissions simultaneously, which can further reduces the transmission latency. Furthermore, MEC server can serve multiple users simultaneously based on NOMA technology, which can improve the system performance as well.
\subsection{Practical Implementation Considerations}
In practice, reducing complexity and overhead is an important issue for the design of FD NOMA networks. For example, resource allocation schemes based on perfect CSI can provide a better performance than the schemes based on limited CSI feedback. Nevertheless, an optimal or near-optimal channel estimation algorithms involves severe communication overhead and relative high computational complexity. Therefore, resource allocation schemes for FD NOMA systems relying on partial CSI should be investigated to reduce the required control signal exchange.

Moreover, FD NOMA networks raise higher requirement on SI cancellation capability. When receiving the superposed signals, the FD device should perform SIC procedure in the presence of SI. Once the SI cannot be mitigated effectively, the SIC procedure cannot be successful. For this reason, the capability of SI cancellation has a great impact on SIC procedure. Consequently, well-designed SI mitigation schemes are necessary for performing SIC procedure.

In addition, the accuracy of SIC technology plays a key role in the practical system performance. Once the messages of previous users are not correctly decoded by SIC, decoding error will accumulate for the remaining users. However, most of the existing works rely on the assumption that SIC can perfectly canceling the interference. Therefore, research on FD NOMA under the assumption of imperfect SIC is a promising research direction in the future.

\section{Conclusions\label{sect: conclusion}}
In this article, the application of FD technology to NOMA is surveyed and studied. Several typical transmission protocols incorporating these two technologies are presented. Then, a novel FD NOMA system based on C-RAN architecture is proposed, where the RRHs operate DL and UL NOMA in FD mode, and the baseband processing is carried out at the CU. Furthermore, power allocation policies are discussed for this scenario, and simulation results are provided to demonstrate the superiority of this scheme. Finally, challenges and research opportunities of FD NOMA systems are identified to stimulate the future research.
\bibliography{NOMAFDR}
\begin{IEEEbiography}
{Xianhao Chen}  (xianhaochen@ufl.edu) is currently a Ph.D. student with the Department of Electrical and Computer Engineering, University of Florida. He received the B.Eng. degree in communication engineering from Southwest Jiaotong University in 2017. His research interests include vehicular networks, edge computing, and non-orthogonal multiple access.
\end{IEEEbiography}
\begin{IEEEbiography}
{Gang Liu} [M¡¯15] (gangliu@swjtu.edu.cn) is currently a lecturer at the School of Information Science and Technology, Southwest Jiaotong University (SWJTU), Chengdu, China. He received the Ph.D. degree in Communication and Information Systems from Beijing University of Posts and Telecommunications (BUPT) in 2015. His current research interests include next generation wireless networks, massive machine-type communications, full-duplex wireless and resource management. Dr. Liu has co-authored more than 30 technical papers in international journals and conference proceedings. He won the Excellent Doctoral Dissertation Award of BUPT in 2015 and the Best Paper Award in IEEE ICC'2014. He is now serving as the secretary and treasurer for IEEE ComSoc, Chengdu Chapter.
\end{IEEEbiography}
\begin{IEEEbiography}
{Zheng Ma} [M¡¯07] (zma@home.swjtu.edu.cn) is currently a professor at Southwest Jiaotong University, and serves as deputy
dean of the School of Information Science and Technology. His research interests include information theory and coding,
signal design and applications, FPGA/DSP Implementation, and professional mobile radio (PMR). He has published more than
60 research papers in high-quality journals and conferences. He is currently an Editor for IEEE Communications Letters. He is also the Chairman of the Communications Chapter of the IEEE Chengdu section.
\end{IEEEbiography}
\begin{IEEEbiography}{Xi Zhang} (S'89-SM'98-F'16)
Xi Zhang [F] received his B.S. and M.S. degrees from Xidian University, Xi¡¯an, China, and an M.S. degree from Lehigh University,Bethlehem, Pennsylvania, all in electrical engineering and computer science, and his Ph.D. degree in electrical engineering and computer science (electrical engineering-systems) from the University of Michigan, Ann Arbor. He is currently a full professor and the founding director of the Networking and Information Systems Laboratory, Department of Electrical and Computer Engineering, Texas A\&M University, College Station. He is a Fellow of the IEEE for contributions to quality of service (QoS) theory in mobile wireless networks. He is an IEEE Distinguished Lecturer of both IEEE Communications Society and IEEE Vehicular Technology Society. He also received also received a TEES Select Young Faculty Award for Excellence in Research Performance from the Dwight Look College of Engineering at Texas A\&M University, College Station, in 2006. He was with the Networks and Distributed Systems Research Department, AT\&T Bell Laboratories, Murray Hill, New Jersey, and AT\&T Laboratories Research, Florham Park, New Jersey, in 1997. He was a research fellow with the School of Electrical Engineering, University of Technology, Sydney, Australia, and the Department of Electrical and Computer Engineering, James Cook University, Australia. He has published more than 320 research papers on wireless networks and communications systems, network protocol design and modeling, statistical communications, random signal processing, information theory, and control theory and systems. He received the U.S. National Science Foundation CAREER Award in 2004 for his research in the areas of mobile wireless and multicast networking and systems. He received Best Paper Awards at IEEE ICC 2018, IEEE GLOBECOM 2014, IEEE GLOBECOM 2009, IEEE GLOBECOM 2007, and IEEE WCNC 2010, respectively. One of his IEEE Journal on Selected Areas in Communications papers has been listed as the IEEE Best Readings Paper (receiving the highest citation rate among all IEEE transactions/journal papers in the area) on Wireless Cognitive Radio Networks and Statistical QoS Provisioning over Mobile Wireless Networking. He also received a TEES Select Young Faculty Award for Excellence in Research Performance from the Dwight Look College of Engineering at Texas A\&M University, College Station, in 2006. He is serving or has served as an Editor for IEEE Transactions on Communications, IEEE Transactions on Wireless Communications, IEEE Transactions on Vehicular Technology, and IEEE Transactions on Network Science and Engineering, twice as a Guest Editor for the IEEE Journal on Selected Areas in Communications for two Special Issues, one on Broadband Wireless Communications for High Speed Vehicles and the other on Wireless Video Transmissions,
an Associate Editor for IEEE Communications Letters, twice as Lead Guest Editor for IEEE Communications Magazine for
two Feature Topics, one on Advances in Cooperative Wireless Networking and the other on Underwater Wireless Communications
and Networks: Theory and Applications, and a Guest Editor for IEEE Wireless Communications for a Special Issue on Next Generation CDMA vs. OFDMA for 4G Wireless Applications, an Editor for Wiley¡¯s Journal on Wireless Communications and Mobile Computing, the Journal of Computer Systems, Networking, and Communications, and Wiley¡¯s Journal on Security and Communications Networks, and an Area Editor for Elsevier¡¯s Journal on Computer Communications, among many others. He is serving or has served as the TPC Chair for IEEE GLOBECOM 2011, TPC Vice-Chair for IEEE INFOCOM 2010, TPC Area Chair for IEEE INFOCOM 2012, Panel/Demo/Poster Chair for ACM MobiCom 2011, General Chair for IEEE WCNC 2013, TPC Chair for IEEE INFOCOM 2017¨C2018 Workshops on Integrating Edge Computing, Caching, and Offloading in Next Generation Networks, and TPC/General Chair for numerous other IEEE/ACM conferences, symposia, and workshops.
\end{IEEEbiography}
\begin{IEEEbiography}
{Pingzhi Fan} [M¡¯93-SM¡¯99-F¡¯15] (pzfan@home.swjtu.edu.cn) received his Ph.D. degree in electronic engineering from Hull
University, United Kingdom. He is currently a professor and director of the Institute of Mobile Communications, Southwest
Jiaotong University. His research interests include high mobility wireless communications, machine learning in wireless networks, signal design, and coding.
\end{IEEEbiography}
\begin{IEEEbiography}
{Shanzhi Chen} [SM¡¯04] (chensz@datanggroup.cn) received his Ph.D. degree from Beijing University of Posts and Telecommunications(BUPT), China, in 1997. He joined Datang Telecom Technology $\&$ Industry Group in 1994, and has
served as CTO since 2008. He was a member of the steering expert group on information technology of the 863
Program of China from1999 to 2011, and received Outstanding Young Researcher Award from the Nature Science
Foundation of China in 2014. He is the director of the State Key Laboratory of Wireless Mobile Communications,
and a board member of Semiconductor Manufacturing International Corporation (SMIC). He has devoted his work
to the research and development of TD-SCDMA 3G and TD-LTE-Advanced 4G since 2004. He received the State Science
and Technology Progress Award in 2001 and 2012. His current research interests include network architectures,
wireless mobile communications, the Internet of Things, and vehicular network.
\end{IEEEbiography}
\begin{IEEEbiography}
{Fei Richard Yu} [S¡¯00, M¡¯04, SM¡¯08, F¡¯18] (Richard.Yu@carleton.
ca) is a professor at Carleton University, Canada. His research
interests include connected vehicles, security, and wireless.
He serves on the editorial boards of several journals, including
Co-Editor-in-Chief for Ad Hoc $\&$ Sensor Wireless Networks, lead
series editor for IEEE Transactions on Vehicular Technology, and
IEEE Transactions on Green Communications and Networking,
and IEEE Communications Surveys $\&$ Tutorials. He is a Distinguished
Lecturer and the Vice President (Membership) of the
IEEE Vehicular Technology Society.
\end{IEEEbiography}
\end{spacing}
\end{document}